\documentclass[9pt,twoside]{pnas-new}

\templatetype{pnasmathematics} 
\usepackage{siunitx}
\usepackage{booktabs}
\usepackage{caption}
\usepackage{subcaption}

\title{On the suitability of hBN as an insulator for 2D~material-based ultrascaled CMOS devices}

\author[a]{Theresia~Knobloch}
\author[a,b]{Yury~Yu.~Illarionov}
\author[c]{Fabian~Ducry}
\author[a]{Christian~Schleich}
\author[d]{Stefan~Wachter}
\author[d]{Thomas~Müller}
\author[a]{Michael~Waltl}
\author[e]{Mario~Lanza}
\author[b]{Mikhail~I.~Vexler}
\author[c]{Mathieu~Luisier}
\author[a]{Tibor~Grasser}

\affil[a]{Institute for Microelectronics, TU Wien, Gusshausstrasse 27-29, 1040 Vienna, Austria }
\affil[b]{Ioffe Institute, Polytechnicheskaya 26, 194021 St-Petersburg, Russia }
\affil[c]{Integrated Systems Laboratory, ETH Zürich, Gloriastrasse 35, 8092 Zürich, Switzerland}
\affil[d]{Institute for Photonics, TU Wien, Gusshausstrasse 27-29, 1040 Vienna, Austria }
\affil[e]{Institute for Functional Nano \& Soft Materials, Soochow University, 199 Ren-Ai Road, Suzhou 215123, China}
\leadauthor{T. Knobloch} 

\begin{abstract}
Complementary metal oxide semiconductor (CMOS) logic circuits at the ultimate scaling limit place the utmost demands on the properties of all materials involved. 
The requirements for semiconductors are well explored and could possibly be satisfied by a number of layered two-dimensional (2D) materials, like for example transition-metal dichalcogenides or black phosphorus. The requirements for the gate insulator are arguably even more challenging and difficult to meet.
In particular the combination of insulator to semiconductor which forms the central element of the metal oxide semiconductor field effect transistor (MOSFET) has to be of superior quality in order to build competitive devices. 
At the moment, hexagonal boron nitride (hBN) is the most common two-dimensional insulator and widely considered to be the most promising gate insulator in nanoscaled 2D~material-based transistors. Here, we critically assess the material parameters of hBN and conclude that while its properties render hBN an ideal candidate for many applications in 2D nanoelectronics, hBN is most likely not suitable as a gate insulator for ultrascaled CMOS devices.  
\end{abstract}

\begin{document}
\maketitle
\thispagestyle{firststyle}
\ifthenelse{\boolean{shortarticle}}{\ifthenelse{\boolean{singlecolumn}}{\abscontentformatted}{\abscontent}}{}
\vspace*{-1.8cm}

\section{Introduction} \label{sec:intro} 


Research activities on 2D~material-based nanoelectronics have focused mainly on semiconductors\cite{Chhowalla2016}. While semiconducting 2D materials might be able to outperform silicon at the ultimate scaling limit\cite{Akinwande2019}, for FETs with dimensions scaled down to few atomic layers, suitable insulators are required. 
To maximize the flexibility offered by van der Waals heterostructures, these insulators ideally should provide a layered 2D structure\cite{Geim2013}.
The 2D insulator which is currently widely considered the most promising is hBN\cite{Liu2016a}.
While numerous studies have analyzed the potential and benefits of hBN when used as a substrate and gate insulator of transistors at the proof-of-concept level where thick insulators are employed\cite{Dean2010a}\cite{Zhang2017b}\cite{Rhodes2019}, there is only a limited number of studies about the suitability of hBN as an insulator at the scaling limit of 1-6 atomic layers thickness (\SI{0.33}{nm}-\SI{2}{nm})\cite{Britnell2012a}\cite{Ji2016}\cite{Chen2020}. 
A commonly held view is that this lack is caused by the immaturity of this novel material system and the technological difficulties related to device fabrication based on only a few layers of hBN.

In this perspective we investigate the performance limits of hBN in its ultra-thin form.
First, we summarize the current state of the art in the synthesis of hBN to clearly demonstrate where the main technological challenges lie for batch fabrication 
of few-layer structures to be used as part of the gate stacks in nanoscaled FETs. One of the main benefits of including hBN in the gate stack is the perfectly clean van der Waals interface it forms with other 2D materials. 
This is an advantage over conventional 3D insulators like SiO\textsubscript{2} or HfO\textsubscript{2} which typically exhibit large densities of dangling bonds and charged impurities at the interfaces with 2D materials. 
These interfacial defects act as scattering centers and, in addition to causing other problems, severely degrade the mobility.
However, like any other material, hBN is not free of atomic defects, and films grown using different methods show very different defect types and densities. 
Most importantly for FETs, these defects 
significantly increase the leakage currents through thin hBN layers via trap assisted tunneling (TAT).

We will argue here that, the comparatively high tunnel currents through thin hBN layers make it difficult for hBN to satisfy the stringent requirements for scaled insulators. 
Insulators in modern FETs should be thinner than \SI{1}{nm} equivalent oxide thickness (EOT), which means that less than \SI{1}{nm} of SiO\textsubscript{2} would give the same electrostatic control over the channel.
Due to the low dielectric constant of hBN ($\varepsilon \sim5$) this corresponds to thin hBN films of less than 4 layers(\SI{1.32}{nm}). 
This issue is analyzed by studying experimental and theoretical tunnel currents through hBN.
Calculations are performed for perfectly single-crystalline, defect-free hBN and therefore give a lower estimate for the best case 
leakage current levels through thin hBN layers. 
We will demonstrate that even in the most optimistic case hBN is unlikely a good choice for a gate insulator in nanoscaled 2D CMOS logic.

\section{Synthesis of hBN} \label{sec:synthesis}


At the current stage it is difficult to identify the scaling potential of hBN from experiments, since high-quality, single crystalline multilayer films are difficult to fabricate and currently altogether impossible to grow with batch-compatible processing, as requested by industry standards. At the same time, strong variations in sample properties of hBN films are clearly not yet competitive with mainstream Si technology which has been optimized in a tremendous industrial effort for over half a century. 
Therefore, in order to better understand whether the aforementioned limitations of hBN are a consequence of its immature processing or of its intrinsic material parameters, the current state-of-the-art of hBN synthesis is briefly summarized first.   

A large number of studies use hBN flakes exfoliated from high-purity crystals. 
Naturally, the quality of the single crystals from which few layer samples are exfoliated is key as it directly determines the purity and crystal quality of the hBN samples after exfoliation. 
As for any high-purity material, the growth of high-quality single crystalline hBN is very challenging. Currently the most successful method for growing hBN single crystals was developed by Taniguchi et al.\cite{Taniguchi2007} and relies on high pressures of around \SI{5}{GPa} and high temperatures of up to \SI{1650}{\degreeCelsius}.
Equally important as a highly pure crystal to start from is the careful execution of the mechanical exfoliation itself. Soon after the demonstration of mechanical exfoliation using  adhesive tape to thin down and isolate monolayer graphene, the method was applied to obtain mono- and few-layer samples of hBN\cite{Pacil2008}.  
However, mechanical exfoliation is an inherently random process where single ``good'' flakes have to be selected out of hundreds of ``bad'' flakes that do not meet the requirements regarding thickness homogeneity, size and shape.
Therefore, this method is incompatible with the processing standards required by industrial applications which is why scalable batch processes for the fabrication of hBN layers are heavily investigated. 

The most widely explored scalable growth method is chemical vapor deposition (CVD), where hBN can be deposited in a furnace under a wide range of low pressures of \SI{0.01}{Pa} to \SI{100}{Pa} at temperatures in between \SI{1000}{\degreeCelsius} and \SI{1300}{\degreeCelsius} from the reaction of ammonia borane or borazine on a copper surface\cite{Kim2012a}. 
Recently, single crystalline large area hBN monolayers have been grown on sputter-deposited and annealed crystalline Cu (111) surfaces on c-plane sapphire wafers, where hBN orients itself during growth along the Cu (111) steps\cite{Chen2020}. 
However, one monolayer of hBN is too thin (\SI{0.33}{nm}) to act as insulator in FETs as it allows for high leakage currents, which is why the growth of single crystalline multilayer hBN is required. 
The growth of multilayer hBN is even more challenging as the reaction can no longer be catalyzed by the copper substrate.
Jang et al.\cite{Jang2016} overcame this limitation by using an excess partial pressure of the borazine precursor which, however, resulted in a polycrystalline layer. 
A different approach to avoid the surface-mediated, self-limited growth of multilayer hBN on copper is to use a different substrate such as iron which possesses high solubility for boron and nitrogen and can be used to mediate a precipitation growth mechanism\cite{Kim2015b}.
Shi et al.\cite{Shi2020} developed this approach further by using a liquid Fe-B alloy as a catalyst for epitaxial precipitation of hBN on (0001) sapphire substrates. In their vapor-liquid-solid growth approach, 
they successfully demonstrated the synthesis of
highly crystalline multilayer hBN samples on large areas of up to $\SI{10}{cm^2}$.    

In addition to CVD, three other important growth techniques potentially offer epitaxial growth of single-crystalline, multilayer hBN on a large scale, namely metal-organic chemical vapor deposition (MOCVD), molecular beam epitaxy (MBE) and atomic layer deposition (ALD). MOCVD is based on the same process as CVD but uses gaseous metal-organic compounds as precursors, like for example triethylborane and ammonia for hBN growth. 
Reasonable crystal quality at moderate growth rates can be obtained when growing hBN on c-axis sapphire\cite{Li2016d}.
Contrary to CVD which relies on the catalytic effect of the substrate, MBE  allows for direct in-situ growth of vertically stacked heterostructures\cite{Elias2019}. 
ALD growth relies on gaseous precursors, for example boron trichloride and ammonia, which are introduced in alternating pulses into the growth chamber where material is deposited on a heated substrate\cite{Lee2020}.
The growth temperature for this process is at \SI{600}{\degreeCelsius} still too high for good compatibility with standard CMOS processes but much lower than in all other studies where the growth temperature exceeds \SI{1000}{\degreeCelsius}.

While all four methods of depositing hBN
have seen tremendous progress over the last decade, the growth of single-crystalline multilayered hBN on large areas remains highly challenging. Therefore, the exploration of the benefit of single crystalline bulk hBN as a substrate in nanoelectronics currently still relies on the use of exfoliated flakes as a convenient test platform.


\section{Improved electronic transport for 2D semiconductors in hBN based heterostructures} \label{sec:improvement}


The main advantage of hBN as gate insulator is the clean van der Waals interface it forms with 2D semiconductors. 
In fact, the ultimate thinness of 2D semiconductors makes electronic transport through these layers particularly susceptible to the impact of interfaces and their surroundings in general.
Scanning tunneling spectroscopy measurements have demonstrated that not only the topography of 2D semiconductors is determined by the roughness of the underlying SiO\textsubscript{2} substrate but also the charge density and the band gap vary dramatically depending on charges trapped in the SiO\textsubscript{2}\cite{Rhodes2019}. If instead of SiO\textsubscript{2} hBN is used as a substrate, the atomically flat van der Waals interface substantially reduces charge disorder in the system and minimizes extrinsic sources of charge carrier scattering such as surface roughness as well as scattering by charged impurities and remote phonons\cite{Wang2013b}.

By performing temperature dependent measurements of the mobility, two main sources of scattering can be identified and isolated, namely phonon scattering that dominates at temperatures $>\SI{100}{K}$ and impurity scattering that determines the resistivity at low temperatures. 
Cui et al.\cite{Cui2015} have shown that the low-temperature, impurity dominated mobility in MoS\textsubscript{2} samples is up to two orders of magnitude higher when encapsulated with 10-\SI{30}{nm} thick hBN layers than when placed directly on SiO\textsubscript{2}. This is caused by the reduced defect density in hBN as compared to the SiO\textsubscript{2} interface.
For scattering at these interfacial defects, two different mechanisms can be distinguished, long-ranged Coulomb scattering at charged impurities in the insulator and scattering at short-ranged, interfacial imperfections such as dangling bonds.

The interfacial densities of charged impurities can be extracted by calculating the scattering rates at interfacial Coulomb potentials. In this way, the low-$T$ mobility of hBN encapsulated MoS\textsubscript{2} was modeled with a charged impurity concentration of \SI{6e9}{cm^{-2}}\cite{Cui2015} which is more than two orders of magnitude lower than the concentration of fixed interfacial charges of \SI{e12}{cm^{-2}} observed in MoS\textsubscript{2} on SiO\textsubscript{2}\cite{Baugher2013}.
The Coulomb scattering model also shows that an increased distance of the charge centroid, located in the center of the semiconducting layer, from the Coulomb scattering centers results in a mobility enhancement. This can be realized either through an increase of the thickness of the semiconductor or through the introduction of an hBN interlayer in between the semiconductor and the conventional amorphous oxide.
It was shown that the mobility improves by one order of magnitude if the MoS\textsubscript{2} thickness is increased by \SI{2}{nm}\cite{Cui2015} and it is expected that an hBN interlayer would also have to be at least \SI{3.3}{nm} thick (corresponding to 10 hBN layers) in order to be able to effectively screen fixed charges at the underlying interface with SiO\textsubscript{2}. 

The dielectric environment does not only influence the impurity-dominated low-temperature mobility in 2D semiconductors but also the phonon-limited mobility at room temperature. Even electrons in ultrathin semiconductor layers can excite phonons in the surrounding dielectrics via long-ranged Coulomb interactions, if the dielectric supports polar vibrational modes\cite{Ma2014}. This so-called ``remote phonon'' or ``surface optical'' (SO) phonon scattering has been identified as the dominant phonon scattering mechanism in graphene devices\cite{Konar2010}.
If the dielectric interface has a high density of interfacial charged impurities, e.g. above \SI{e12}{cm^{-2}}, Coulomb scattering at these defects dominates. As a consequence, the mobility is higher if dielectrics with large dielectric constants (high-k dielectrics) such as for example HfO\textsubscript{2} are used because the enhanced dielectric screening reduces Coulomb scattering.
If, on the other hand, the interface is clean and has a small density of charged impurities, SO phonon scattering dominates and the mobility is degraded the most if the 2D layer is surrounded by high-k dielectrics, as they allow low-energy polar vibrational modes\cite{Ma2014}. 
In theory, it is possible to form heterostructures of a high-k dielectric and a hBN layer at the interface to the 2D semiconductor which would improve the mobility without sacrificing the high permittivity of the overall stack. However, one might expect that in order to efficiently damp out the SO phonon scattering several nanometers of hBN would be required, even though this has to be investigated further.

\begin{figure}[!ht]
\centering
\includegraphics[width=\linewidth]{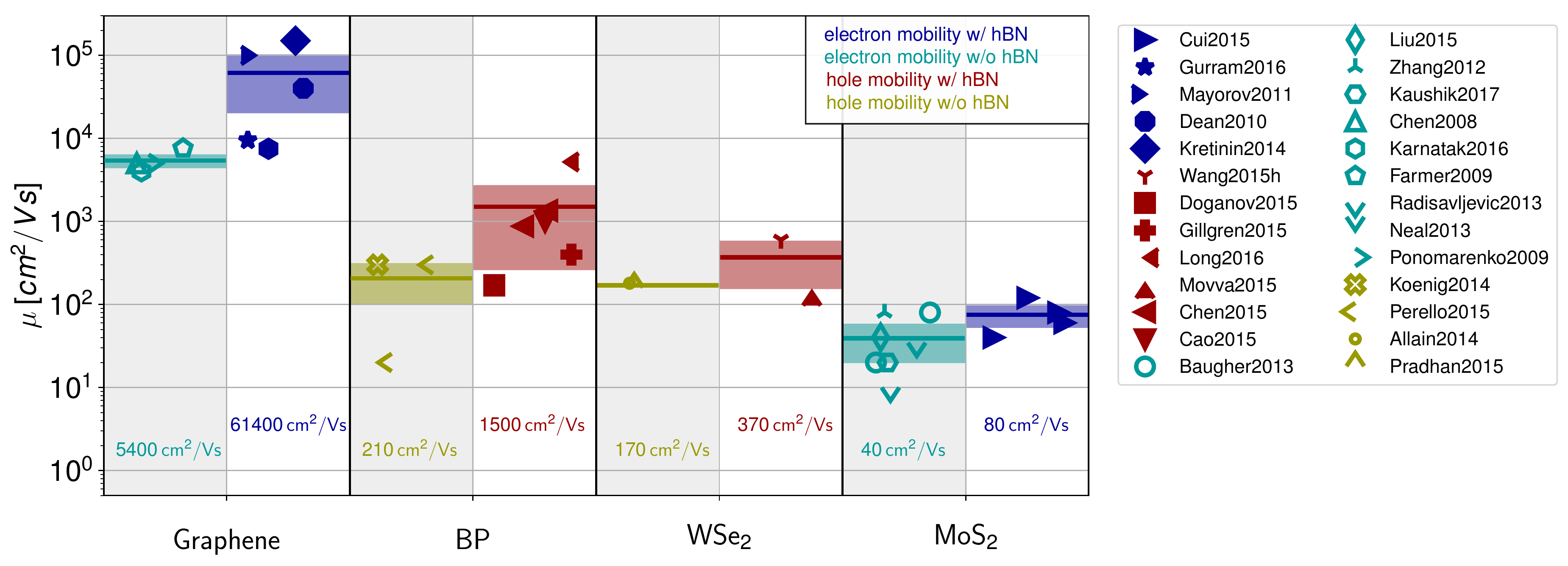}
 \caption{\textbf{Mobility in hBN heterostructures.} For 2D materials the mobility increases if hBN forms a direct interface with the 2D semiconductor. The mobility values are collected from \cite{Cui2015, Gurram2016, Mayorov2011, Dean2010, Kretinin2014, Wang2015h,Doganov2015, Gillgren2015, Long2016, Movva2015, Chen2015, Cao2015, Baugher2013, Liu2015a, Zhang2012, Kaushik2017, Chen2008, Karnatak2016, Farmer2009, Radisavljevic2013, Neal2013, Ponomarenko2009,  Koenig2014, Perello2015, Allain2014a, Pradhan2015a}, where four-probe measurements, ideally Hall measurements, were performed, thereby ensuring the accuracy of the values.
 }
 \label{fig:mob}
\end{figure}

A considerably improved mobility in FETs with a gate stack that includes high-quality multilayer hBN at the interface to the 2D channel material has been indeed demonstrated in numerous experiments as can be seen in Fig. \ref{fig:mob}, where the mobility of MOSFETs with and without an hBN interlayer is compared. In order to ensure the highest possible accuracy of the given values, only four-probe measurements of the mobility and ideally Hall mobility measurements at room temperature on devices fabricated from exfoliated layers are taken into account.
The prospect of maintaining a high mobility ($>$ \SI{100}{cm^2/Vs}) is the key advantage of 2D materials at the ultimate scaling limit over ultra-thin silicon layers\cite{Akinwande2019}, thus a high mobility is essential.

In addition to an improved mobility, hBN enhances device performance in other ways as well. For instance,
the reduced scattering in hBN based heterostructures decreases the inhomogeneous broadening of the excitonic linewidth\cite{Cadiz2017}.
Furthermore, novel physical phenomena, like for example superconductivity in magic angle graphene\cite{Cao2018c}, rely on an encapsulation in high-quality hBN multilayers. 
Beyond its use as a substrate, hBN is a promising material on its own for numerous applications.
Capacitors based on multilayer hBN show a pronounced resistive switching behavior which can be used in resistive random access memories(RRAM) which in turn could serve as solid-state synapses in neuromorphic circuits\cite{Shi2018a}. The small dielectric constant of hBN can be decreased further if it is deposited in amorphous form, thereby rendering it an ideal candidate for an interconnect isolation material\cite{Hong2020}. 

\section{Limitations of hBN as gate insulator for 2D~material-based CMOS} \label{sec:limitations} 

\begin{figure}[!ht]
\centering
\includegraphics[width=.65\linewidth]{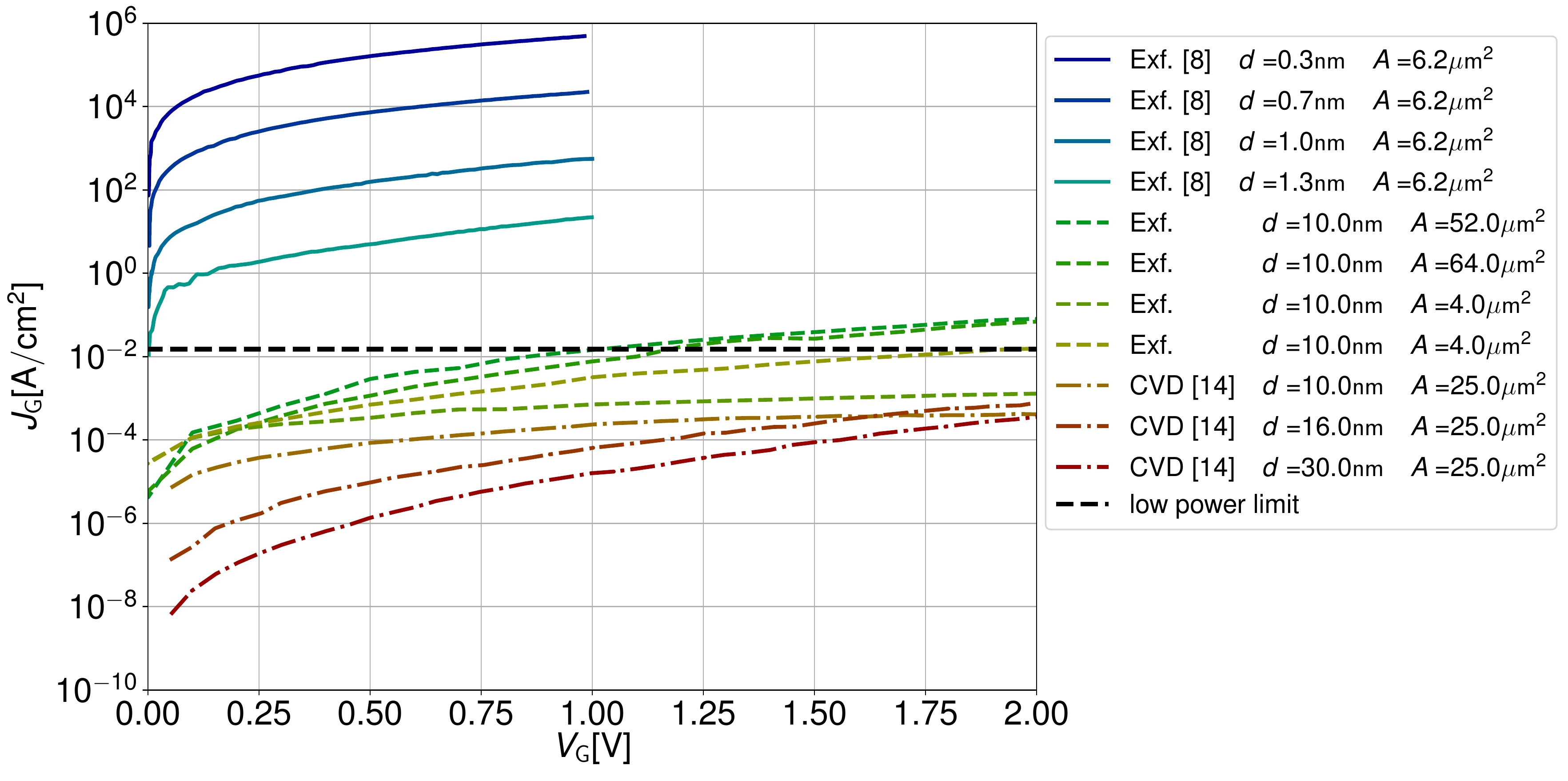}
 \caption{\textbf{Measured leakage currents through hBN.} A comparison of the experimental tunnel current densities through hBN as reported in literature\cite{Britnell2012a}\cite{Jang2016} and measured on our samples. Every line corresponds to a different device. Some of the devices are based on exfoliated (Exf.) hBN while others rely on CVD grown hBN, marked by dash-dotted lines. Dashed lines are for exfoliated samples of EOT$>$8nm, solid lines for EOT$\leq$1nm.
  }

 \label{fig:tunexp}
\end{figure}

Unlike optoelectronic devices and radio-frequency FETs, digital CMOS logic FETs require insulators with sub-\SI{1}{nm} EOT to be competitive with ultra-scaled silicon technologies. 
Moreover, these devices operate at high electric fields and thus could benefit from the high dielectric strength of hBN\cite{Palumbo2019}. 
However, hBN has a moderate band gap ($E_\mathrm{G}\sim\SI{6}{eV}$) and  a comparatively small dielectric constant ($\varepsilon \sim5$) which requires the use of thin layers as, for example, only 3 atomic layers correspond to an EOT of \SI{0.76}{nm}. 
In consequence, the measured tunnel leakage currents through hBN are high ($J \sim \SI{e2}{A/cm^{2}}$ for $\left|V_\mathrm{G}\right| \leq \SI{0.7}{V}$) \cite{Britnell2012a}. 
What is more, at the current state-of-the-art, the reported currents through hBN typically vary over orders of magnitude, see Fig. \ref{fig:tunexp}.
This suggests that the leakage current strongly depends on the material quality and thus on the growth method used. 
The dependence on the defect density in hBN shows that the leakage current is dominated by TAT\cite{Chandni2015a}. 
Such large leakage currents would in turn render hBN unsuitable as gate insulator of ultra-scaled CMOS technologies, as they would considerably increase the off-state currents of the FET. 
In the following, the origins of the large tunnel currents are discussed to evaluate whether these currents are purely defect-mediated and could potentially be sufficiently reduced in hBN samples of very high quality.

Localized defect states within the band gap of hBN give rise to conductive channels in the hBN and as such the density of defects and consequently the density of conductive channels determines the current density.
Therefore, in order to give a realistic estimate of the tunnel current density through a hBN layer it is very important to know the prevalent defect types and corresponding defect densities in the sample. 
In principle point defects, local aggregates of point defects and line defects like dislocations are distinguished. Every point defect is characterized by its vertical depth within the hBN stack, its energetic trap level within the band gap of hBN and its atomic structure. 
The defect types can be characterized from two perspectives, from purely theoretical calculations based on density functional theory (DFT)\cite{Weston2018} or from defect spectroscopy experiments on hBN samples\cite{Greenaway2018}. 
Weston et al.\cite{Weston2018} performed first-principles calculations on native point defects and atomic impurities in hBN. 
They found that atomic carbon, oxygen and hydrogen impurities in hBN have comparatively small formation energies and are expected to be present in significant concentrations in undoped hBN. Strand et al.\cite{Strand2020} analyzed divacancies and found that boron and nitrogen vacancies form stable configurations where adjacent layers are connected with inter-layer molecular bridges. 
One might expect that these inter-layer connections play an important role in current transport through the hBN layers in vertical direction. 

Greenaway et al.\cite{Greenaway2018} studied resonant tunneling through localized states within an hBN tunnel barrier between two graphene electrodes at low temperatures of a few Kelvin. 
From the analysis of tunneling via hBN defect states an increased defect density in between adjacent hBN layers was found, thereby confirming the hypothesis that
defect clusters where molecular bridges connect adjacent layers play an important role for tunneling processes through hBN.
Other experimental methods to analyze the defect density in hBN include high-resolution transmission electron microscopy on freestanding hBN layers\cite{Jin2009}
or scanning tunneling microscopy using graphene as a capping layer to create a conductive surface\cite{Wong2015}.
However, all experimental methods to explore defect states in hBN layers reported so far have focused on identifying single atomic defect states which is not sufficient for a comprehensive modeling of the impact of prevalent traps on the tunnel current. 
As recently large scale synthesis of high quality hBN samples has been successfully demonstrated\cite{Shi2020}, these breakthroughs open up the path towards performing non-destructive analysis of the defect densities throughout the whole band gap of hBN using capacitance voltage or charge pumping measurements where large device areas are required.
In addition, most experimental defect analyses in hBN have focused so far on exfoliated hBN of the highest quality and only few report on hBN grown using other methods. In fact, a reference comparing the logarithmic tunnel current density as a function of gate voltage for samples with given hBN thicknesses and electrode areas could provide valuable information about different defect densities in hBN synthesized with different methods.
A comparison of available current densities is shown in Fig. \ref{fig:tunexp} which may serve as a blueprint for future comparisons of differently synthesized hBN.

\section{Performance projections for scaled hBN} \label{sec:projections}

\subsection{Insulator performance}
In general, a good gate insulator for FETs is characterized by three main properties. First, an insulator has to show a high permeability to electric fields ($\varepsilon >8$) to ensure a good gate control of the surface potential in the channel by the applied gate voltage. This is linked to the main driving force for the reduction of the insulator thickness $d$ in scaled devices as otherwise the gate would lose control over a channel of reduced length, a phenomenon commonly termed short-channel effects. 
Second, the gate leakage current has to be small ($J < \SI{e-2}{A/cm^{2}}$ for $\left|V_\mathrm{G}\right| \leq \SI{0.7}{V}$\cite{IRDS2020}) to reduce the off-state current of the transistor and thereby the stand-by power consumption of the device. Third, the defect density at the interface and in the insulator should be as small as possible ($N < \SI{e10}{cm^{-2}}$) to maintain a high mobility in the semiconductor and to enhance device stability. 

For devices based on 2D materials, there is an additional fourth requirement for the gate insulator, which is to provide an as high room-temperature mobility in the semiconductor as possible (ideally above $\SI{200}{cm^2/Vs}$) by minimizing remote phonon scattering. As discussed above, few-nanometer hBN is ideally suited for this purpose. However, hBN layers with a thickness of a few nanometer lead together with the rather small dielectric constant of hBN to a reduced gate control. The ideal value for the dielectric constant of the gate insulator in 2D material FETs is governed by a fundamental trade-off, as high values cause good gate control and small gate leakage currents but small values, like in hBN, assure high mobilities. 

\subsection{Methods}
To estimate the performance potential of hBN, we want to establish a theoretical lower limit of how small tunnel leakage currents through hBN could be if there were no defects within the layer.
For this purpose we calculated the tunnel current through hBN based on the intrinsic material properties of ideal, single crystalline, multilayer hBN. 
According to these theoretical considerations the tunnel current densities through hBN could be significantly smaller than the values experimentally reported, possibly even staying below the low power limit ($J = \SI{1.5e-2}{A/cm^{2}}$ for $\left|V_\mathrm{G}\right| \leq \SI{0.7}{V}$).
In the following analysis we will consider this most optimistic scenario and focus on the TAT-free best case.

\begin{figure}
 \begin{subfigure}[c]{0.42\textwidth}
        \includegraphics[width=\textwidth]{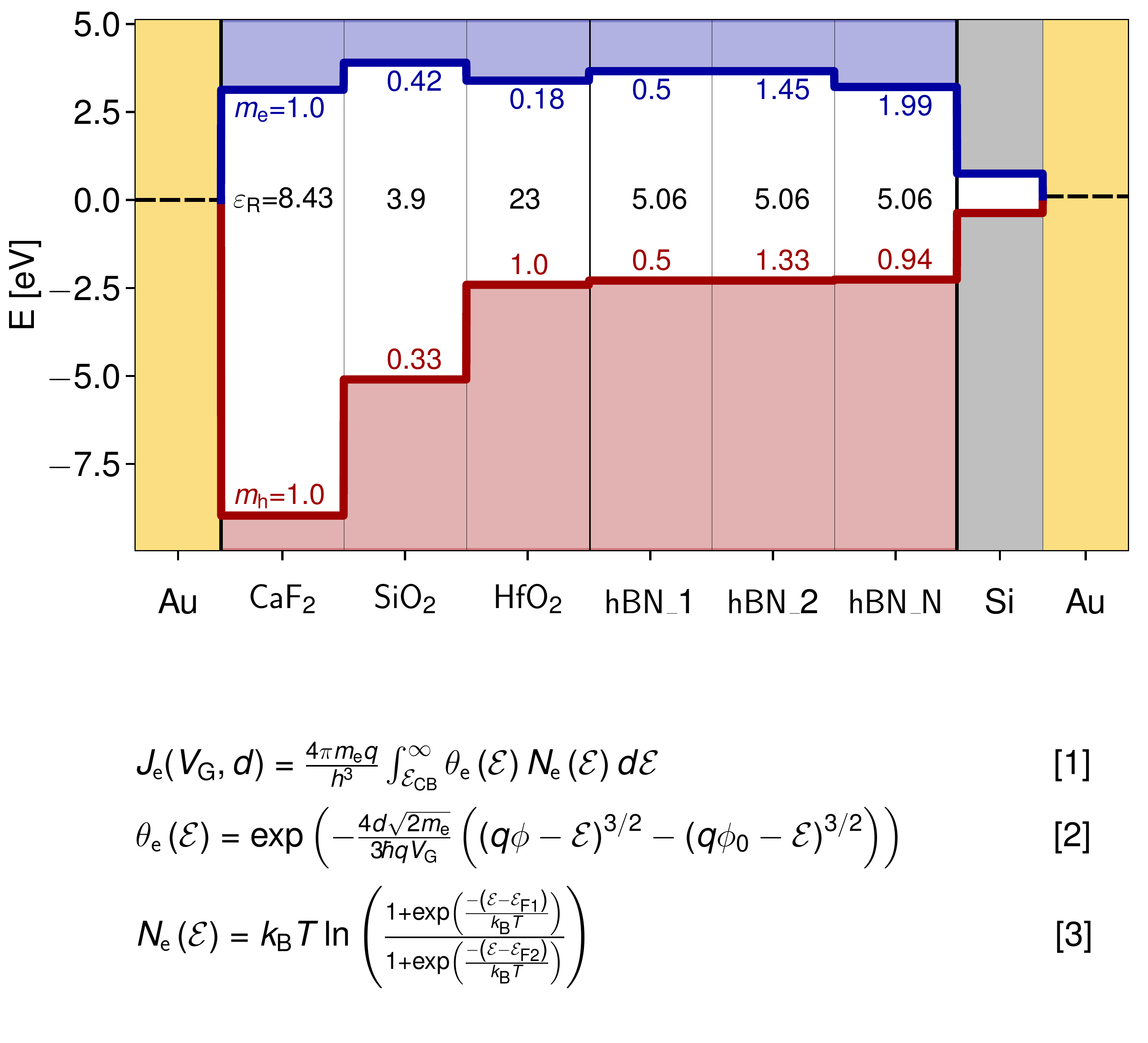}
        \caption{Band diagram and Tsu-Esaki model.}
        \label{fig:band}
    \end{subfigure}
    \hspace*{0.25cm} 
    \begin{subfigure}[c]{0.4\textwidth}
     \begin{tabular}{@{}c|lll|c|c|cc@{}}
 \toprule
  Trend & \multicolumn{3}{l|}{Parameter}& Value & Reference& Used for \\
  \midrule
 $\uparrow$ & bandgap & $E_\mathrm{G}$ & & \SI{5.95}{eV} & \cite{Arnaud2006}\cite{Cassabois2016} & hBN\_1, \_2\\ 
     &   &   & & \SI{5.472}{eV} & - & hBN(NEGF)\\
 \midrule
 - & electron affinity & $\chi$ & & \SI{1.14}{eV} & \cite{Haastrup2018} & hBN\_1,  \_2  \\
      &  &  & & \SI{1.3}{eV} & \cite{Strand2020} &  - \\
  &  &  & & \SI{1.59}{eV} & - &  hBN(NEGF) \\
 \midrule
  $\uparrow$ &dielectric const. & $\varepsilon$ & & $5.06\:\varepsilon_0$ & \cite{Geick1966} & hBN\_1, \_2\\
  \midrule
  $\uparrow$ &electron mass & $m_\mathrm{e}$ & - & $0.5\:m_0$ & \cite{Britnell2012a} & hBN\_1\\
  & & & - & $1.45\:m_0$ & - & hBN\_2\\
  & & &  (M$\rightarrow$L)& $2.21\:m_0$ & \cite{Xu1991} & -\\
   & & &  ($\Gamma\rightarrow$A)& $0.26\:m_0$ & - &  hBN(NEGF)\\
  & & &  (M$\rightarrow$L)& $67.6\:m_0$ & - &  hBN(NEGF)\\
  & & &  (H$\rightarrow$K)& $0.94\:m_0$ & - &  hBN(NEGF)\\
  \midrule
  $\uparrow$ &hole mass & $m_\mathrm{h}$ & -  & $0.5\:m_0$ & \cite{Britnell2012a} & hBN\_1 \\
  & & & (M$\rightarrow$L)& $1.33\:m_0$ & \cite{Xu1991} & hBN\_2\\
  & & &  ($\Gamma\rightarrow$A)& $20.1\:m_0$ & - &  hBN(NEGF)\\
  & & &  (M$\rightarrow$L)& $1.45\:m_0$ & - &  hBN(NEGF)\\
  & & &  (H$\rightarrow$K)& $1.99\:m_0$ & - &  hBN(NEGF)\\

  \bottomrule
\end{tabular} 
\caption{hBN material parameters.}
\label{tab:params}
    \end{subfigure}\\
    
\caption{
 \textbf{Material parameters of hBN.}
 In Figure \ref{fig:band} the Tsu-Esaki equations describe the electron contribution to the direct tunnel current density $\mathsf{J_\mathsf{e}}$ through an insulating barrier. 
 In addition, the band alignment of hBN is shown in comparison with other common dielectrics and several sets of plausible material parameters of hBN are given.
 The material parameters of hBN are listed in Table \ref{tab:params}. In the leftmost column of the table to the right, the trends for suppressing tunnel leakage currents are highlighted, where $\uparrow$ stands for as high as possible and $-$ for not specifiable in general. 
}
\end{figure}

We performed simulations for a system of 3 layers of hBN corresponding to an EOT of \SI{0.76}{nm} according to the current technology node\cite{IRDS2020}. The hBN layers were placed in between a gold electrode with a work function of \SI{4.7}{eV} and a bottom p-doped silicon layer ($N_\mathsf{A}=\SI{e18}{cm^{-3}}$, $N_\mathsf{D}=\SI{e10}{cm^{-3}}$), thereby forming a MIS structure. 
The current through this structure was calculated using the Tsu-Esaki model\cite{Tsu1973} as implemented in Comphy\cite{Rzepa2018}. 
Within this model the tunneling transmission probability is approximated by a Wentzel-Kramers-Brillouin (WKB) factor and the expressions for the electron current are given in Equations [1] - [3] in Fig. \ref{fig:band}. Similar expressions are also valid for the hole current, i.e. for carrier transport between the Si valence band and the metal. 
The tunnel current depends most strongly on the parameters in the exponential WKB factor, given in Equation [2], and as such on the layer thickness ($d$), the applied voltage ($V_\mathsf{G}$) and on the material parameters of the energy barrier heights ($\phi, \phi_\mathsf{0}$) and the effective tunnel masses for electrons ($m_\mathsf{e}$) and holes ($m_\mathsf{h}$). Another material parameter that indirectly affects the layer thickness is the dielectric constant which defines the equivalent oxide thicknesses ($\mathrm{EOT} = \frac{\mathcal{\epsilon}_\mathsf{R}}{\mathcal{\epsilon}_\mathsf{R, SiO_2}} d_\mathsf{SiO_2}$).
The band diagrams for hBN according to various plausible sets of material parameters are shown in Fig. \ref{fig:band} in comparison to other commonly used insulators. In addition, in Table \ref{tab:params} the material parameters of hBN are explicitly listed and the parameter trends required for efficiently suppressing a direct tunnel current are included. For minimizing the tunnel current, the band gap has to be large to ensure high energy barriers, the effective masses have to be large and the dielectric constant also has to be as high as possible, as this corresponds to a large physical layer thickness for a given EOT, even though this contradicts the endeavor of high charge carrier mobilities in the semiconductor.

It is important to note that with this methodology we can only give a range of tunnel current estimates through hBN, shown as a blue shaded area in Fig. \ref{fig:perf_curr}. This is because the effective tunnel masses for electrons and holes are important but empirical parameters that have only barely been studied in the past. Here, we used two sets of parameters.
The first set of small tunnel masses, corresponding to high currents, was taken from Ref.\cite{Britnell2012a}, where the tunnel masses were calibrated to experimental data. We hypothesize that the agreement achieved in this work was likely due to a severe underestimation of the effective tunnel masses that control the tunnel current which compensated for the neglect of TAT in the model. This set was used for the simulations of $\mathrm{hBN\_\mathrm{1}}$. The second set of high tunnel masses, corresponding to small currents, was adapted from Ref.\cite{Xu1991}, where the effective masses were extracted from DFT calculations of the band structure of bulk hBN. 
This set was used for the simulations of $\mathrm{hBN\_\mathrm{2}}$. The calculation of the effective tunnel masses for out of plane transport through hBN based on the DFT band structure is nearly impossible for two main reasons.
First of all, two valleys contribute to the out of plane transport in bulk hBN (the M$\rightarrow$L and H$\rightarrow$K valleys). 
Second, the band structure of hBN along these two orientations is nearly flat which corresponds to high tunnel masses but also leads to large uncertainties in the extraction of the effective masses from bands with a small curvature. This small curvature directly originates from the layered structure of hBN and as such from the high inherent anisotropy of a layered material. Therefore, the effective electron mass is most likely overestimated in Ref.\cite{Xu1991} and was adapted here to a smaller, more plausible value based on the currents we simulated with \textit{ab initio} methods, as explained below. The wide blue shaded area in Fig. \ref{fig:perf_curr} that spans in some regions more than 4 orders of magnitude demonstrates the importance of future studies on extracting effective tunnel masses for electrons and holes through hBN.

In order to avoid the problem of calculating effective masses and to increase the accuracy of the estimated currents through defect-free hBN, we simulated the current through Au-hBN-Si structures using a non-equilibrium Green's functions (NEGF) approach combined with DFT. 
In these full-band transport simulations the Hamiltonian and overlap matrices of the considered devices were first calculated with the CP2K package\cite{Kuhne2020} based on a hybrid functional\cite{Heyd2003}. These matrices were then loaded into a quantum transport solver\cite{Bruck2017} to obtain the current-voltage characteristics of the MIS stack using the same electrostatic potentials as in the WKB case. 

\subsection{Performance projections}
\begin{figure}[!ht]
\centering
    \begin{subfigure}[c]{0.55\textwidth}
        \includegraphics[width=\textwidth]{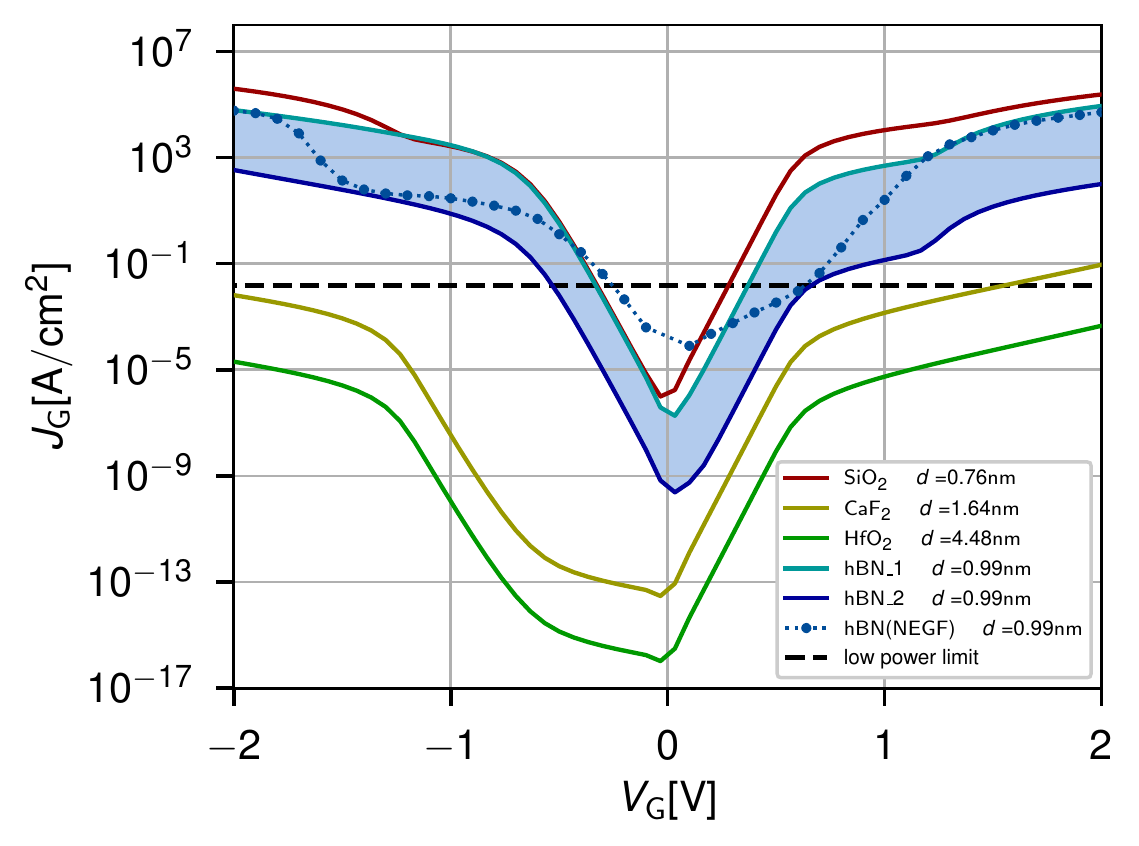}
        \caption{Gate leakage currents for $\mathrm{EOT}=\SI{0.76}{nm}$.}
        \label{fig:perf_curr}
    \end{subfigure}\\
    \begin{subfigure}[c]{0.49\textwidth}
        \includegraphics[width=\textwidth]{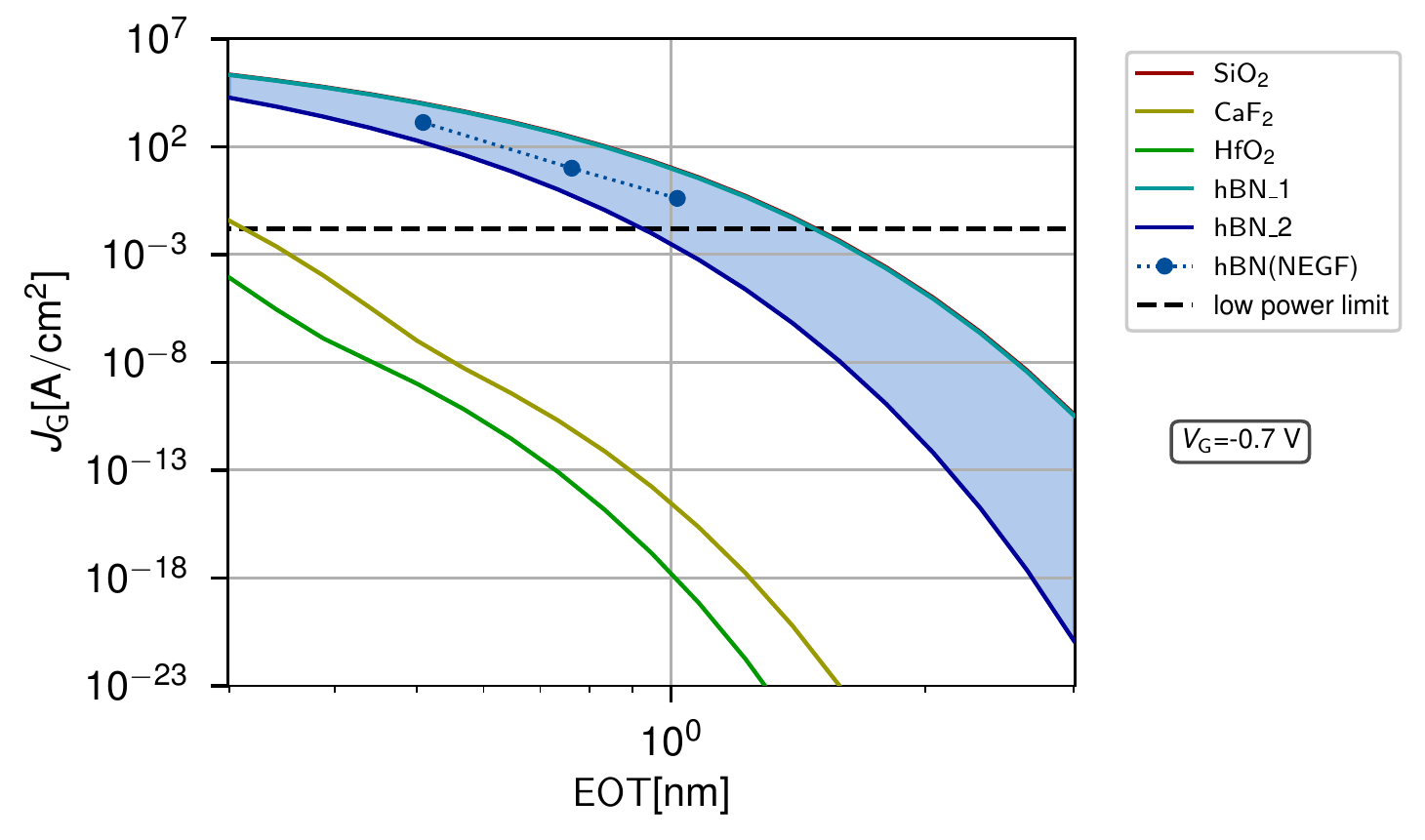}
        \caption{Gate leakage currents for a pMOS at a constant gate bias. \\ $V_\mathsf{G}=\SI{-0.7}{V}$.}
        \label{fig:J_EOT_V_pMOS}
    \end{subfigure}
    \begin{subfigure}[c]{0.49\textwidth}
        \includegraphics[width=\textwidth]{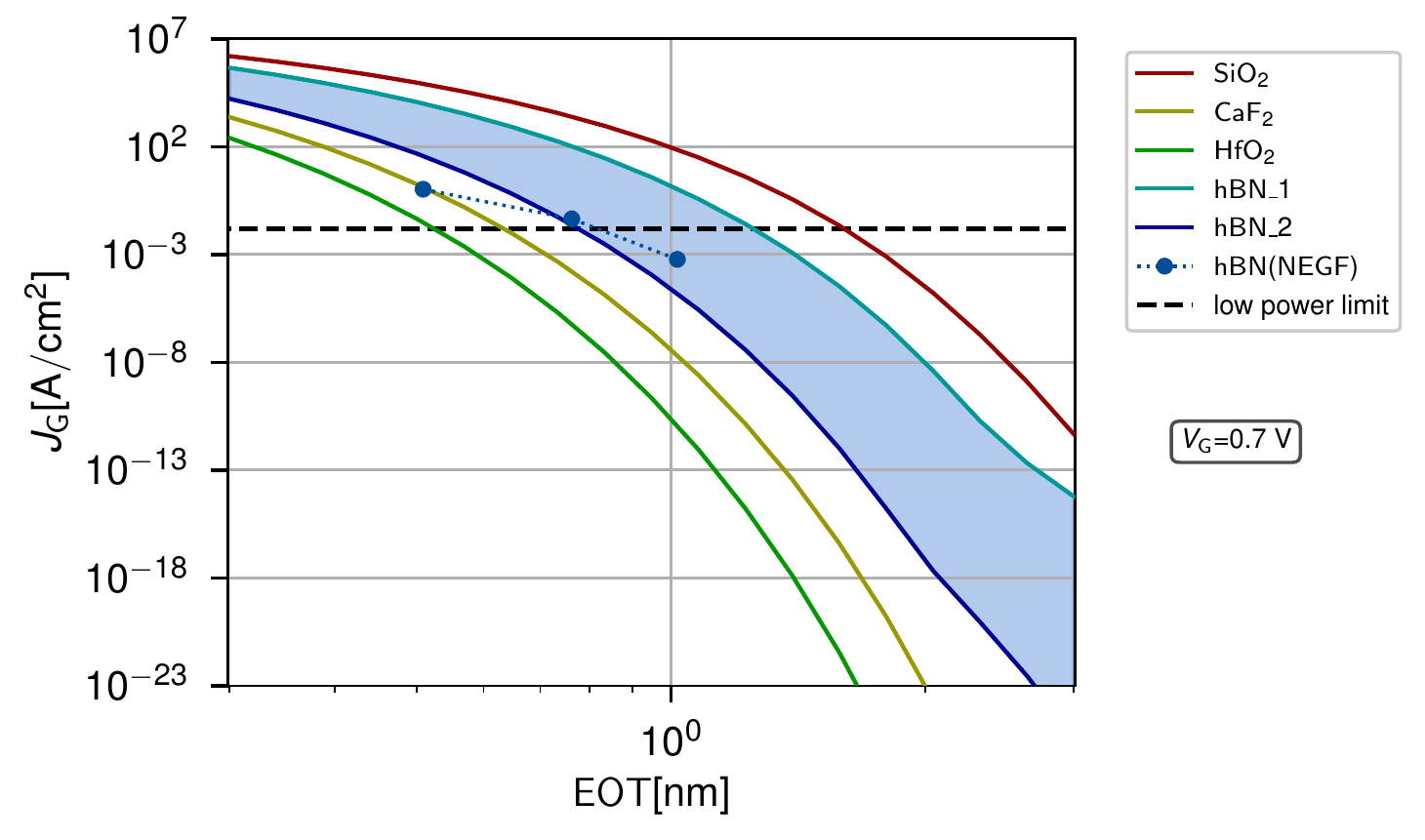}
        \caption{Gate leakage currents for a nMOS at a constant gate bias. \\  $V_\mathsf{G}=\SI{0.7}{V}$.}
        \label{fig:J_EOT_V_nMOS}
    \end{subfigure} \\
    \begin{subfigure}[c]{0.49\textwidth}
        \includegraphics[width=\textwidth]{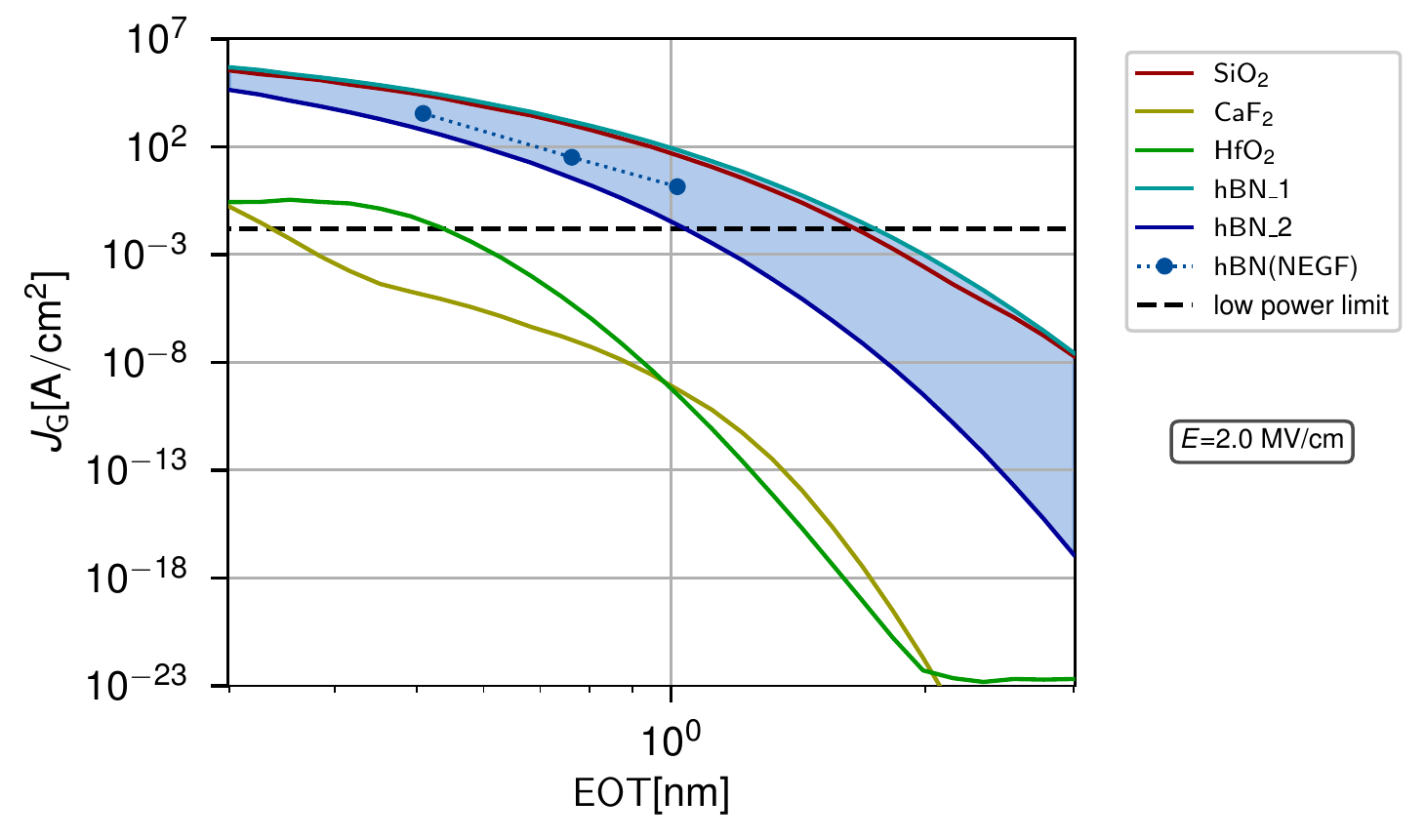}
        \caption{Gate leakage currents for a pMOS at a constant electric field. \\ $V_\mathsf{G}<\SI{0}{V}$.}
        \label{fig:J_EOT_E_pMOS}
    \end{subfigure}
    \begin{subfigure}[c]{0.49\textwidth}
        \includegraphics[width=\textwidth]{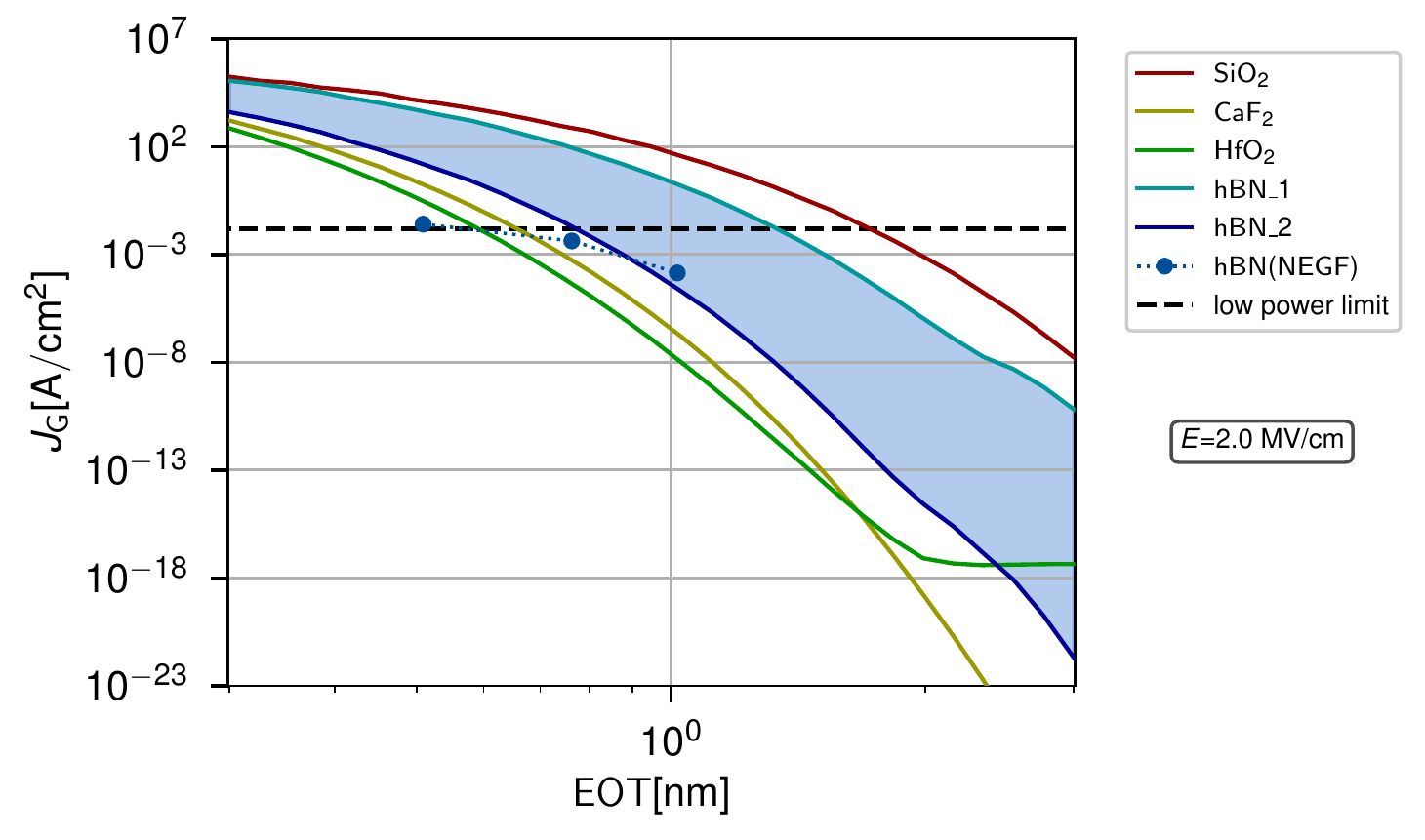}
        \caption{Gate leakage currents for a nMOS at a constant electric field. \\ $V_\mathsf{G}>\SI{0}{V}$.}
        \label{fig:J_EOT_E_nMOS}
    \end{subfigure}
    
 \caption{ \textbf{Performance projection of the tunnel current through hBN in the defect-free case.} A comparison of the tunnel current densities through hBN and other dielectrics based on the Tsu-Esaki model for the MOS stack of Au/dielectric/Si is shown for insulators with a thickness corresponding to an EOT of \SI{0.7}{nm} in Figure \ref{fig:perf_curr}. In addition,  \textit{ab initio} data is included as full circles. 
 The current density as a function of the EOT is compared for different dielectrics at a constant gate voltage in Figures \ref{fig:J_EOT_V_pMOS} and \ref{fig:J_EOT_V_nMOS} and at a constant electric field of \SI{2}{MV/cm} in Figures \ref{fig:J_EOT_E_pMOS} and \ref{fig:J_EOT_E_nMOS}.
 }
 \label{fig:perf}
\end{figure}

The results of both approaches (DFT+NEGF and the Tsu-Esaki range) are in good agreement, as can be seen in Fig. \ref{fig:perf_curr}.
Only for small gate voltages the tunnel current is underestimated by the range we obtain from the Tsu-Esaki model.
When comparing the leakage currents through hBN with other insulators it becomes clear that at a small EOT of \SI{0.76}{nm} the gate leakage through hBN is slightly lower than through SiO\textsubscript{2}. However, both are orders of magnitude higher than through the high-k dielectric hafnium dioxide (HfO\textsubscript{2}) and through crystalline calcium fluoride (CaF\textsubscript{2})\cite{Wen2020}.
In addition, we calculated the tunnel current as a function of EOT for a fixed electric gate field and a fixed applied voltage, corresponding to the two scaling laws for avoiding short channel effects when decreasing device dimensions, Dennard scaling and constant voltage scaling.
The tunnel currents as a function of the EOT are shown for a negative voltage of $\SI{-0.7}{V}$ in Fig. \ref{fig:J_EOT_V_pMOS} and for a positive voltage of $\SI{+0.7}{V}$ in Fig. \ref{fig:J_EOT_V_nMOS}, corresponding to a pMOS and a nMOS of the 2020 technology node, where the supply voltage amounts to $V_\mathrm{DD}=\SI{0.7}{V}$\cite{IRDS2020}. In Fig. \ref{fig:J_EOT_V_pMOS} it can be clearly seen that hBN is not suitable as a gate insulator in a pMOS for scaled devices with EOT $<$\SI{1}{nm}, as the gate leakage current exceeds the low-power limit by more than one order of magnitude. If the leakage currents are above the low-power limit, the off-state currents and thus the power consumption of the device are too high for e.g. applications in consumer electronics. This observation was made for the current technology node and will be even worse for future nodes.
The conclusions are the same when comparing the tunnel current density as a function of EOT for constant electric gate fields of \SI{2}{MV/cm}, as shown in Fig. \ref{fig:J_EOT_E_pMOS} for negative voltages (pMOS) and in Fig. \ref{fig:J_EOT_E_nMOS} for positive voltages (nMOS). There are substantial leakage currents even through defect-free hBN when negative voltages are applied due to the small band offset for holes of only about $\phi_\mathsf{h}=\SI{1.9}{eV}$ and the comparatively small tunnel masses for holes in the M$\rightarrow$L and H$\rightarrow$K valleys of hBN. 

These comparisons show that there is in general a shortage of insulators which provide sufficient leakage current blocking potential for continued down-scaling of devices and insulators. Among all insulators currently available, hBN is particularly ill-suited for being used as a scaled gate dielectric for pMOS devices and shows a performance slightly worse than most insulators for nMOS devices at operation voltages below \SI{1.0}{V}; above this regime the current through hBN dramatically increases.

One possible solution to this lack of scalable insulators would be to go to an operation regime of small supply voltages where only steep-slope devices can operate. 
In conventional MOSFET devices the subthreshold swing (SS) of a FET cannot be smaller than the Boltzmann limit of 60mV/decade at room temperature. In order to overcome this limit the device operation principles have to be modified. Approaches that could achieve this goal exploit a constrained injection energy window for charge carriers such as Tunnel FETs\cite{Appenzeller2004},
a nonmonotonic variation around the Fermi energy in the density-of-states as an energy filter\cite{Qiu2018a},
or an insulator that creates a negative capacitance of the gate stack like for example a ferroelectric, thereby amplifying the surface potential of the channel\cite{Salahuddin2008}.
All these steep-slope devices allow for a supply voltage below \SI{0.5}{V} where also hBN could serve as a gate insulator. In addition, hBN can also be used for applications where tunneling through the layer is required as part of the device design like a tunneling barrier in a Tunnel FET based on a graphene/hBN heterostructure\cite{Britnell2012}.

\section{Conclusions} \label{sec:conclusions} 


At the moment, hBN is widely considered to be the most promising insulator for FETs based on 2D materials. In the last decade a lot of progress has been made in developing growth methods of hBN compatible with batch processing, but in particular the synthesis of single crystalline, multilayer hBN with small impurity concentrations on large areas has not yet been achieved.
However, in experimental studies based on exfoliated devices and theoretical calculations it was shown that an interface of a 2D semiconductor and more than 5 layers of hBN can provide charge carrier mobilities in the 2D semiconductor higher than for any other insulating layer used so far to form the interface. This can be attributed to the small density of charged impurities that would act as Coulomb scattering centers and to its small dielectric constant which reduces remote phonon scattering. These properties made the observation of novel physical phenomena possible\cite{Dean2013} and make it an attractive material in itself for numerous applications in analogue devices\cite{Gaskell2015} as well as in nano- and optoelectronics\cite{Caldwell2019}.

Nonetheless, the excessive tunnel currents through hBN are at present dominated by defect states in the hBN band gap, such as atomic impurities of carbon, oxygen and hydrogen atoms and 
defect clusters which form molecular bridges between hBN layers. 
Currently, a comprehensive picture about energetic trap levels and defect densities for hBN based on different synthesis methods as well as a benchmark of the defect dominated tunnel currents through layers of different qualities is not available.
Thus, to evaluate the theoretically attainable lower limit of the tunnel current through hBN, we assumed defect-free layers and calculated the tunnel currents using both a Tsu-Esaki and a DFT+NEGF based model. Both methods provided good agreement and showed that while for ultrascaled EOT the leakage current through hBN is slightly lower than through SiO\textsubscript{2}, it is orders of magnitude higher than through HfO\textsubscript{2} or CaF\textsubscript{2} at the same EOT. In particular for pMOS devices with an EOT below \SI{1}{nm} hBN is highly unsuitable as a gate insulator due to its small dielectric constant, its relatively high valence band edge and moderate effective tunnel masses.
Based on the calculated leakage currents we have also shown that there is currently a general lack of insulators compatible with 2D materials which would sufficiently block currents to allow for continued scaling while at the same time maintaining the low power consumption required in consumer electronic devices. The ideal gate insulator would offer large band gaps, high tunnel masses, a moderate dielectric constant, a crystalline structure and a high quality van der Waals interface with 2D materials. We believe that only an intensive search for suitable gate insulators or gate insulator stacks formed by combinations of several insulators together with the adaption of a concept for steep-slope transistors will deliver the performance boost expected by next-generation ultra-scaled CMOS logic.

\section*{References}

\bibliography{refs_new}

\newpage

\section*{Figures}

\end{document}